# Direct Visualization of a Static Incommensurate Antiferromagnetic Order by Suppressing the Superconducting Phase Coherence in Fe-doped Bi$_2$Sr$_2$CaCu$_2$O$_{8+\delta}$

Siyuan Wan[1†], Huazhou Li[1†], Peayush Choubey[2†], Qiangqiang Gu[1†], Han Li[1], Huan Yang[1*], Ilya M. Eremin[2*], G. D. Gu[3] & Hai-Hu Wen[1*]

[1] National Laboratory of Solid State Microstructures and Department of Physics, Center for Superconducting Physics and Materials, Collaborative Innovation Center for Advanced Microstructures, Nanjing University, Nanjing China

[2] Institut für Theoretische Physik III, Ruhr-Universität Bochum, D-44801 Bochum, Germany

[3] Condensed Matter Physics and Materials Science Department, Brookhaven National Laboratory, Upton, New York 11973, USA

†These authors contributed equally to this work. *e-mail: huanyang@nju.edu.cn; Ilya.Eremin@ruhr-uni-bochum.de; hhwen@nju.edu.cn.

**Abstract:** In cuprate superconductors, due to strong electronic correlations, there are multiple intertwined orders which either coexist or compete with superconductivity. Among them the antiferromagnetic (AF) order is the most prominent one. In the region where superconductivity sets in, the long-range AF order is destroyed. Yet the residual short-range AF fluctuations are present up to a much higher doping and their role in the emergence of the superconducting phase is still highly debated. Here, by using a spin polarized scanning tunneling microscope, for the first time, we directly visualize an emergent incommensurate AF order in the nearby region of Fe impurities embedded in the optimally doped Bi$_2$Sr$_2$CaCu$_2$O$_{8+\delta}$ (Bi2212). Remarkably the Fe impurities suppress the superconducting coherence peaks with the gapped feature intact, but pin down the ubiquitous short-range incommensurate AF order. Our work shows an intimate relation between antiferromagnetism and superconductivity.

Since the discovery of cuprate superconductors, the question on the nature of the Cooper pairing mechanism has attracted enormous attention and has been regarded as the most important issue in contemporary condensed matter physics[1]. Due to the strong electronic correlations, various intertwined orders[2-6] are proposed to either compete or co-exist with superconductivity, yet there is no consensus on whether these are foes or friends for realizing high-$T_c$ superconductivity. Now it is widely conceived that the parent phase of the cuprate superconductors is a Mott insulator with a (π, π) antiferromagnetic order. Through slight hole doping to a level of about 0.02 hole/Cu[7], the long-range AF order is destroyed, the insulating behavior becomes suppressed, and superconductivity emerges gradually at the doping of 0.05 hole/Cu. At a doping level of about 0.16 hole/Cu, the superconducting transition temperature reaches a maximum, and this feature seems to be quite unified in different systems[8,9]. At this typical doping level, the AF order does not exist, but some residual AF spin fluctuations are commonly observed in different systems[10-12]. One of the key questions is what role the short-range antiferromagnetic order and its internal structure play and how they link with superconductivity? If the AF spin fluctuation dominates here, one could ask whether the AF order can re-emerge in some particular form if superconductivity is again suppressed. In this study, by using a spin polarized scanning tunneling microscope (SP-STM), for the first time, we directly visualize an incommensurate AF order nearby the magnetic Fe impurities.

On the theory side, we analyze the STM spectra using the single-band Hubbard Hamiltonian with competing *d*-wave superconducting order and short-range antiferromagnetic correlations. We show that the weak-magnetic impurity pins the short-range AF order near the impurity for the interaction strength lower than the critical value of interaction $U_c$, which is necessary to establish long-range antiferromagnetic state in the clean system. We identify the short-range AF order with the incommensurate momentum connecting the nodal regions of the Fermi surface with opposite Fermi velocities as explaining the vast majority of the experimental data. Combination of our experimental and theoretical results shows that magnetic impurities can be used as a powerful tool to elucidate the driving quantum correlations in unconventional superconductors.

The optimally doped (OP) Bi2212 has a critical temperature $T_c$ of about 89 K as seen from resistive and magnetization measurements (Fig. 1a,b), while $T_c$ decreases to about 65 K in the sample with 3% doping of Fe ions to Cu sites. In addition, the Fe substitution broadens the superconducting transition width[13] and increases the normal-state resistivity in the sample. Furthermore, the temperature dependent magnetization measured at 5 T shows a large paramagnetic background above $T_c$ in the Fe doped sample, in contrary to the negligible magnetization background in the OP Bi2212 sample (see Supplementary Fig. S1). From the fitting to the temperature dependent magnetization by using the Curie-Weiss law, the obtained momentum for each Fe site is about 5.0$\mu_B$ with $\mu_B$ the Bohr magneton (see Supplementary Section 1). The value is consistent with a previous report[14], and it confirms that the Fe impurities are magnetic ones. Figure 1c shows a typical STM topographic image with atomic resolution measured on the Fe-doped Bi2212 single crystal by using a tungsten tip. The commonly observed periodic supermodulations of the BiO surface can be seen in the topography, and the square lattice of Bi atoms (with Cu atoms below at exactly the same positions) can be revealed with the lattice constant $a_0$ of about 3.8 Å. In Fig. 1c, there are additional bright spots denoted by arrows on the surface, and these spots are absent on the surface of OP Bi2212 samples[15]. Figure 1d shows the Fourier transformed (FT) pattern of the topographic image, and one can see clear Bragg spots as well as the supermodulation spots. Two tunneling spectra in wide energy window are shown in Fig. 1e, and they are measured at the bright spot site and away from it. A considerable enhancement of the differential conductance can be observed at the bias voltage above 1V on the spectrum measured at the bright spots. This is the reason why the bright spots showing up in the topographic image recorded with the bias voltage of 1.3 V are supposed to be Fe impurities underneath. In Fig. 1f, we show a set of tunneling spectra measured along the black arrowed line crossing an Fe impurity as shown in the inset. At positions far away from the bright spots on the surface, the V-shaped tunneling spectra can be detected with the coherence peak energy of about 40 meV. The spectrum shape and the coherence peak energy are similar to those obtained in the OP Bi2212 sample. On the spectrum measured at the impurity center, one can see that the superconducting coherence peaks are strongly suppressed, indicating the local suppression of superconductivity. It was previously reported that the Fe impurities may introduce low-energy resonances in the density of states[16], which we also witness in the averaged spectra (see Supplementary Section 3 and Fig. S3). This further confirms that the bright spots are induced by Fe impurities. However, we must emphasize that, even with the in-gap bound states at the Fe impurity sites, the intensity of these bound states is much weaker than that induced by Zn impurities[17]. This provides another evidence that the Fe impurities can be regarded as weak magnetic ones.

Figure 2a,b shows the quasiparticle interference (QPI) mappings $g(\mathbf{r}, E)$ measured at different energies ($E$) by using a tungsten tip. The locations of Fe impurities can be identified in QPI mappings especially at $E$ = 60 meV, which is above the superconducting gap due to impurity induced renormalization of the spectral weight (see Supplementary Section 3). The corresponding FT-QPI patterns are shown in Fig. 2d,e. Although the tunneling spectrum is similar to that measured in the OP Bi2212, the seven sets of basic scattering patterns, which connect the octet ends of banana-shaped contour of the constant energy cuts of the quasiparticle energy dispersion[18,19], are missing in the FT-QPI pattern at 20 meV. This may be related to the suppressed spectral weight of the coherent quasiparticles[20,21]. However, there are two well identified sets of characteristic scattering spots in Fig. 2d. These spots have the central wave vectors of $\mathbf{Q'} = (0, \pm\frac{1}{4}\frac{2\pi}{a_0})$, $(\pm\frac{1}{4}\frac{2\pi}{a_0}, 0)$ and $\mathbf{Q''} = (0, \pm\frac{3}{4}\frac{2\pi}{a_0}), (\pm\frac{3}{4}\frac{2\pi}{a_0}, 0)$. The corresponding real-space period is about $4a_0 \times 4a_0$ for the $\mathbf{Q'}$ scattering spots, and it is about $4/3a_0 \times 4/3a_0$ for the $\mathbf{Q''}$ spots. The real-space modulations of $\mathbf{Q'}$ patterns can be distinguished in Fig. 2a, and these were regarded as the checkboard-like electronic modulation in the underdoped cuprates of $Bi_{2-y}Pb_ySr_{2-z}La_zCuO_{6+x}$[22] and $Ca_{2-x}Na_xCuO_2Cl_2$[23], as well as in the vortex cores in OP Bi2212[24,25]. From the data measured at energy of 60 meV, the scattering spots of $\mathbf{Q'}$ disappear, while the intensities of $\mathbf{Q''}$ scattering spots are enhanced. Similar scattering spots of $\mathbf{Q''}$ are also observed in the underdoped $Ca_{1.88}Na_{0.12}CuO_2Cl_2$ and $Bi_2Sr_2Dy_{0.2}Ca_{0.8}Cu_2O_{8+\delta}$[26], which is explained in terms of the density

wave state with a *d*-wave symmetry form-factor[26-28]. Figure 2f plots *q*-dependent FT-QPI intensities along the arrowed lines (in the $q_{11}$ direction), shown in Fig. 2d,e, at different energies. One can see clearly that $4a_0 \times 4a_0$ checkerboard states (**Q'**) exist in the energy range of |*E*| < 30 meV, while the $4/3a_0 \times 4/3a_0$ modulations (**Q"**) have a pronounced amplitude only for *E* > 35 meV. This contrasting behavior indicates that these two orders should have different origins.

    The SP-STM is a very useful tool to investigate the magnetic structures in the nanoscale and also in discerning the chiral spin state[29]. This technique has been successfully applied in visualizing the AF state in $Fe_{1+x}Te$[30]. Since the Fe ions are magnetic impurities and the high-temperature superconductivity is clearly suppressed near them, as shown in Fig. 1, it is very interesting to know what kind of magnetic state is induced nearby the Fe impurities. For this purpose, a Cr tip is used to investigate the physical properties in the samples. It was shown that the body of Cr tip should be in the antiferromagnetic state, but the atom(s) on the very tip should have a magnetic moment and can be polarized by an external magnetic field[31]. This gives an advantage as the tip should give no strong stray magnetic field, but behaves as a magnetic detector through tunneling current. Before doing the SP-STM measurements, we have characterized and checked our tip on the FeTe sample (see Method, Extended Data Fig. 1), which reveals clearly a bi-collinear AF order as observed previously[30]. Figure 3a,b shows the QPI mappings measured at 0 T by using a Cr tip polarized in situ by the magnetic field of +1.2 T (recorded as $g_\uparrow$) and −1.2 T (recorded as $g_\downarrow$), respectively. The Fe impurities again appear as bright spots at the energy of 50 meV. In the corresponding FT-QPI patterns shown in Fig. 3d,e, the feature of $4/3a_0 \times 4/3a_0$ modulations (**Q"**) can still be resolved by the

spin polarized tip. The relatively wider distribution of **Q"** spots may be due to the relatively blunt Cr STM tip as compared with W tip. The QPI mappings of $g_\uparrow$(**r**, 50meV) and $g_\downarrow$(**r**, 50meV) together with the related FT-QPI patterns are very similar to each other, which means that the non-spin-polarized signal is much larger than the spin-polarized one. Hence, we calculate the difference of these two QPI mappings and show the result of spin-difference conductivity $\delta g_{\uparrow-\downarrow}$(**r**, 50 meV) = $g_\uparrow$(**r**, 50 meV) − $g_\downarrow$(**r**, 50 meV) in Fig. 3c. One can clearly observe that there are sign-alternating periodic modulations appearing near Fe impurities, which can be regarded as a newly emergent AF order. In the FT result of Fig. 3c, as shown in Fig. 3f, the $4/3a_0 \times 4/3a_0$ modulations are absent, suggesting that this static order corresponding to the $4/3a_0 \times 4/3a_0$ modulations may not have correlations with the spin structure. Instead, the spin-difference FT-QPI shows four new scattering spots as highlighted by the red circles in Fig. 3f, and their centers are around the wave vector of **Q**$_0$ = $(\pm 0.85 \frac{\pi}{a_0}, \pm 0.85 \frac{\pi}{a_0})$. These scattering patterns correspond very well to the sign-alternating periodic modulations of the spin-difference QPI in the real space, see Fig. 3c. The central directions of the incommensurate modulations in **q**-space are at an oblique angle of 45° comparing to the direction of the Bragg points, and is consistent with the direction of the long-range antiferromagnetic order in the parent compound[32]. However, the length of the wave vector **Q**$_0$ is smaller than that of such long-range antiferromagnetic order $(\pm \frac{\pi}{a_0}, \pm \frac{\pi}{a_0})$. To better show the newly emergent incommensurate AF order, we then conduct the inverse Fourier transform to the areas within the red circles in Fig. 3f, and the result is shown in the inset of Fig. 3d. The sign-alternating periodic modulations of spin-difference QPI are more clearly illustrated in the inset. Our observation seems to be different from that by Lake et al.[33] in $La_{2-x}Sr_xCuO_4$ with x = 0.10. There they observed four emergent scattering spots around the AF wave vector ($\pm\pi$, $\pm\pi$) under a magnetic field of 14.5T, which was interpreted as the occurrence of the possible charge or spin stripes.

A control experiment has been carried out in another smaller area with the dimensions of 6 nm × 6 nm. QPI mappings are carried out at different energies from 10 to 150 meV by the Cr tip polarized by the magnetic field of +1.2 and −1.2 T, respectively (Extended Data Fig. 2). Figure 4a illustrates a representative $\delta g_{\uparrow-\downarrow}(\mathbf{r}, E)$ measured at 50 meV. The sign-alternating periodic modulations are not as pronounced at 10 meV (Extended Data Fig. 2), but they become very clear in the $\delta g_{\uparrow-\downarrow}$ mappings at 50 and 100 meV. The period of the local extrema of the modulations mismatches the Bi atom sites (with Cu atoms below at exactly the same positions) marked by the black dots in real space (Fig. 4a), which confirms the mismatch of the period in **q**-space of the new modulations and Bragg points or long-range antiferromagnetic order in the parent compound. The spatial oscillation of the new modulations can be observed in the line cut of intensity across an Fe impurity (Fig. 4b). The FT result in Fig. 4f also shows four strong scattering spots around characteristic wave vector of $\mathbf{Q}_0 = (\pm 0.85 \frac{\pi}{a_0}, \pm 0.85 \frac{\pi}{a_0})$. Most importantly, the newly observed modulations do not break the $C_4$ rotation symmetry of the crystal. Based on the FT pattern of $\delta g_{\uparrow-\downarrow}(\mathbf{r}, E)$ measured at different energies (Extended Data Fig. 2), the energy dispersion of the new $\mathbf{Q}_0$ modulation is shown in Fig. 4h. The non-dispersive $\mathbf{Q}_0$ modulation is observed with the vector length of about $1.2\pi/a_0$ appearing in the energy range from about 25 meV to 150 meV.

Theoretically, it was predicted that a short range AF order may appear near the strong non-magnetic (Zn) impurities in cuprates[34,35], which appears due to a correlation-induced splitting of an electronic bound state arising from the sign change of the order parameter along quasiparticle trajectories and thus is a direct consequence of the *d*-wave superconductors. However, it was not known till now whether and how the weak magnetic impurities can promote the short- range antiferromagnetic state. In order to interpret our experimental observations, in what follows we analyze this question using microscopic calculations employing a tight-binding model of a *d*-wave superconductor on a square lattice with antiferromagnetic correlations (Details see Supplementary Section 4). We take a weak magnetic impurity and compute spin-resolved continuum LDOS $\rho_\sigma(r,\omega)$ at a height $z \approx$ 4Å above BiO terminated surface in Bi2212. $\rho_\sigma(r,\omega)$ shows LDOS suppression confined in a very small area (radius ~ 2*a*) around the impurity (Extended Data Fig. 3) as expected from a point-like scatterer[36]. However, the spin-difference LDOS $\rho_{\uparrow-\downarrow}(r,\omega)$ (Fig. 4d) shows an extended structure around the impurity (radius ~ 7*a*) with tails extending far from the impurity. In addition, $\rho_{\uparrow-\downarrow}(r,\omega)$ exhibits a strong damping of oscillations in Cu-O direction (Fig. 4e). Fourier transform of $\rho_{\uparrow-\downarrow}(r,\Delta_0)$ is shown in Fig. 4g, and spin-difference QPI $\rho_{\uparrow-\downarrow}(q,\Delta_0)$ shows dominant intensity around scattering vectors between the nodes of the *d*-wave order parameter as illustrated in Fig. 4c. To check the bias-dependence of dominant wave-vectors in spin-difference QPI, we Fourier transform $\rho_{\uparrow-\downarrow}(r,\omega)$ at different energies (Extended Data Fig. 4). We find that as the energy approaches $\Delta_0$, $\rho_{\uparrow-\downarrow}(q,\omega)$ shows strong QPI features around these characteristic scattering vectors, which remain almost dispersionless at larger biases. Fig. 4i plots the dispersion in $\omega$-$q_{11}$ plane, where $q_{11}$ is the wave vector in a nodal direction. We find that for $\omega > 0.15$, $\rho_{\uparrow-\downarrow}(q_{11},\omega)$ shows dispersionless QPI feature at $\mathbf{q}_{11} \approx$

$q_0$, where $q_0 \approx 0.64(2\pi/a)$ is the wave vector joining two nodal points on the Fermi surface, within a large bias range (~ 0.15-0.6). It's worth mentioning that $q_0$ is quite distinct from the hot-spot nesting wavevector $q_h \approx 0.45(2\pi/a)$ for the Fermi surface. Note that a dispersionless feature in spin-difference QPI is in stark contrast with usual spin-sum QPI, which shows wavevectors dispersing within the octet model[36].

To better understand the origin of the dispersionless QPI feature in $\rho_{\uparrow-\downarrow}(q,\omega)$, we studied the effects of changing various parameters such as $\Delta_0$, $U$, and $t'$ in our model; and corresponding results are shown in Extended Data Fig. 5. First, we find that the occurrence of $q_0$ feature is a normal state phenomenon and having a $d$-wave condensate just shifts the energy at which it first appears. As shown in Extended Data Fig. 5a,b and 4i, respectively, the $q_0$ feature appears at $\omega = 0$ for the normal state ($\Delta_0 = 0$), at $\omega \approx 0.05$ when $\Delta_0 = 0.1$, and at $\omega \approx 0.15$ when $\Delta_0 = 0.3$ (all units of the nearest neighbor hopping). Second, having AF correlations is crucial as indicated by the absence of $q_0$ feature for $U = 0$, see Extended Data Fig. 5c. Third, the $q_0$ feature indeed corresponds to the wavevector connecting two nodes along $q_{11}$ direction as demonstrated by changing the Fermi surface (using $t' = -0.4$) and observing a corresponding shift in $q_0$, see Extended Data Fig. 5d.

The agreement between our theory calculations and experimental data points towards the new role of weak magnetic impurities playing in the study of unconventional superconductors. In contrast to the strong non-magnetic impurities, which induce a local AF state in the $d$-wave superconductor due to the sign change of the superconducting order parameter, weak magnetic impurities tend to pin an intrinsic AF instability of the system leading to the formation of an incommensurate AF order with the wave vectors connecting the nodal points at the

Fermi surface. Although the consistency between our experiment and theoretical picture is quite encouraging, we mention that further theoretical studies would be necessary to elucidate the interplay of d-wave superconducting and antiferromagnetic orders in the presence of magnetic impurities. Our observation of an incommensurate AF order nearby Fe impurities in Fe-doped OP Bi2212 certainly shed new light towards understanding the interplay between superconductivity and AF magnetism and intimately to the pairing mechanism of cuprate superconductors.

## Methods

### Sample and tip preparation

The STM/STS measurements were done on a scanning tunneling microscope (USM-1300, Unisoku Co., Ltd.) with the ultrahigh vacuum up to $10^{-10}$ torr, at low temperatures down to 350 mK, and magnetic fields up to 11 T. A typical lock-in technique is used with an AC modulation of 987.5 Hz. All data are taken at 1.5K. During the experiment, we used two kinds of electrochemically etched tips: conventional tungsten tip and spin-polarized chromium tip. The Cr tip we used was etched by the 3mol/L NaOH solution with the immersed end covering with a length of polytetrafluoroethylene (PTEF) tubing. After the etching, the Cr tip was directly transferred into the STM chamber with ultrahigh vacuum. This as-etched Cr tip was treated and characterized on the surface of the bi-collinear antiferromagnet $Fe_{1+x}Te$. In order to check the spin polarization of the tip, the topographic images $T_\uparrow(\mathbf{r}, +1.2T)$ and $T_\downarrow(\mathbf{r}, -1.2T)$ are measured in the same area of $Fe_{1+x}Te$ under the magnetic field of +1.2T and -1.2T, respectively. Fourier transforms of these topographic images show the Bragg points and antiferromagnetic (AF) peaks located at 1/2 $Q_{Bragg}$, corresponding to a $2a_0$ real-space modulation. Next we obtain the difference of the topographic images $\delta T_{\uparrow-\downarrow}(\mathbf{r}) = T_{\uparrow}(\mathbf{r}, +1.2T) - T_{\downarrow}(\mathbf{r}, -1.2T)$. The background lattice information is subtracted out and only the spin-resolved signals remain. The Te lattice Bragg peaks almost diminish to zero and the antiferromagnetic peaks become dominant. Due to the hysteresis of the magnetic moment on the tip, before the spin polarized measurements, we usually increase the magnetic field to a high value and then reduce the field to zero or a smaller value with the same vorticity. In this way we confirmed that the Cr tip can change its spin-polarization direction by applying the initial magnetic field with different vorticities.

**Theoretical modeling**

We analyze the interplay of short-range AF order and $d$-wave superconductivity, as observed in our experiment, by using microscopic calculations employing a tight-binding model of a $d$-wave superconductor on a square lattice with antiferromagnetic correlations described by the following mean-field Hamiltonian.

$$H_0^{MF} = -\sum_{(i,j),\sigma} t_{ij}\left(c_{i\sigma}^\dagger c_{j\sigma} + h.c.\right) + \sum_{i,\sigma}(Un_{i,-\sigma} - \mu)c_{i\sigma}^\dagger c_{i\sigma} + \sum_{<i,j>}\left(\Delta_{ij}c_{i\uparrow}^\dagger c_{i\downarrow}^\dagger + h.c.\right)$$

Here, in the first term, $<i,j>$ and $(i,j)$ represent only NN, and both NN and NNN sites, respectively, with NN hopping $t_{ij} = t$ that sets the energy scale, and NNN hopping $t_{ij} = t'$. Further, $c_{i\sigma}^\dagger$ ( $c_{i\sigma}$) creates (annihilates) an electron at site $i$ with spin $\sigma$. The second term describes repulsive Hubbard-type interaction, which is mean-field decoupled in the spin density wave channel, producing a Neel-type AF state when the interaction strength $U$ exceeds critical value $U_c$. The last term describes a $d$-wave superconducting condensate with pair field defined on NN bonds as $\Delta_{ij} = V\langle c_{j\downarrow} c_{i\uparrow}\rangle$, where $V$ is the NN attraction. The chemical potential $\mu$ sets the total electron/hole count $n_{tot} = \sum_{i,\sigma} n_{i,\sigma} = \sum_{i,\sigma} \langle c_{i\sigma}^\dagger c_{i\sigma}\rangle$, which we fix at 0.85 per site, corresponding to 15% hole doping. We choose the parameters such that a paramagnetic superconducting state is realized ($U < U_c$). Next, we model a magnetic impurity by a classical spin at site $i^*$ interacting with conduction electrons via local exchange coupling of strength $J$.

$$H_{imp}^{mag} = J \sum_{i^*,\sigma} \sigma c_{i\sigma}^\dagger c_{i\sigma}$$

To study the effects of a single weak magnetic impurity embedded in a *d*-wave superconductor with AF correlations, we diagonalize the total Hamiltonian $H = H_0^{MF} + H_{imp}^{mag}$ using spin-resolved Bogoliubov transformation[37], see the Supplementary Section 4 for the details. The ensuing Bogoliubov-de Gennes (BdG) equations are solved self-consistently and the solution is used to construct lattice Green's functions $G_{ij\sigma}(\omega)$, which yields spin-resolved lattice LDOS at site $i$ and energy $\omega$ as $N_{i\sigma}(\omega) = -\frac{1}{\pi} Im[G_{ii\sigma}(\omega)]$, where $Im$ denotes imaginary part. Note that the differential conductance measured by STM is proportional to the continuum LDOS $\rho(\boldsymbol{r},\omega)$ evaluated at the tip position $\boldsymbol{r}$[38], which can be qualitatively different from the lattice LDOS, especially in presence of inhomogeneity, due to filtering effects caused by intervening charge-reservoir layers between the tip and CuO$_2$ plane. Accordingly, we compute continuum Green's function $G_\sigma(\boldsymbol{r},\omega)$ via the basis transformation from lattice to continuum space as follows.

$$G_\sigma(\boldsymbol{r},\omega) = \sum_{ij} G_{ij\sigma}(\omega) W_i(\boldsymbol{r}) W_j^*(\boldsymbol{r})$$

Here, matrix element of the transformation $W_i(\boldsymbol{r})$ is the Wannier function centered at site $i$. We use first-principles Cu-$d_{x2-y2}$ Wannier functions obtained by downfolding density functional theory (DFT) derived bands to Cu-$d_{x2-y2}$ orbital in Bi2212[39]. Spin-resolved LDOS at the STM tip position will be given by the imaginary part of the continuum Greens function: $\rho_\sigma(\boldsymbol{r},\omega) = -\frac{1}{\pi}Im[G_\sigma(\boldsymbol{r},\omega)]$. Further, we can obtain spin-resolved quasi particle interference (QPI) $\rho_\sigma(\boldsymbol{q},\omega)$ due to the magnetic impurity simply by taking the Fourier transform of $\rho_\sigma(\boldsymbol{r},\omega)$. Experimentally measurable spin-sum and spin-difference QPI can be obtained as $\rho_{\uparrow\pm\downarrow}(\boldsymbol{q},\omega) = \rho_\uparrow(\boldsymbol{q},\omega) \pm \rho_\downarrow(\boldsymbol{q},\omega)$.

**Author contributions:**

The STM/STS measurements and analysis were performed by S.W., H.L., Q.G. H.L., H.Y. and H.-H.W. The samples were grown by G.G. The theoretical calculation was done by P.C. and I.M.E.. S.W., H.Y., H.H.W. and I.M.E. contributed to the writing of the paper. All authors have discussed the results and the interpretations.

**Acknowledgments:** We are grateful to Roland Wiesendanger for the kind advices of how to prepare the spin polarized tip for STM measurements. We also acknowledge Steven A. Kivelson and Joerg Schmallian for useful discussions. This work was supported by National Natural Science Foundation of China (Grants: NSFC-DFG12061131001 (ER 463/14-1), 11927809, 11974171). The work at BNL was supported by the US Department of Energy, office of Basic Energy Sciences, contract no. DOE-sc0012704.

**Competing interests:** The Authors declare no competing interests.

**Figures and legends**

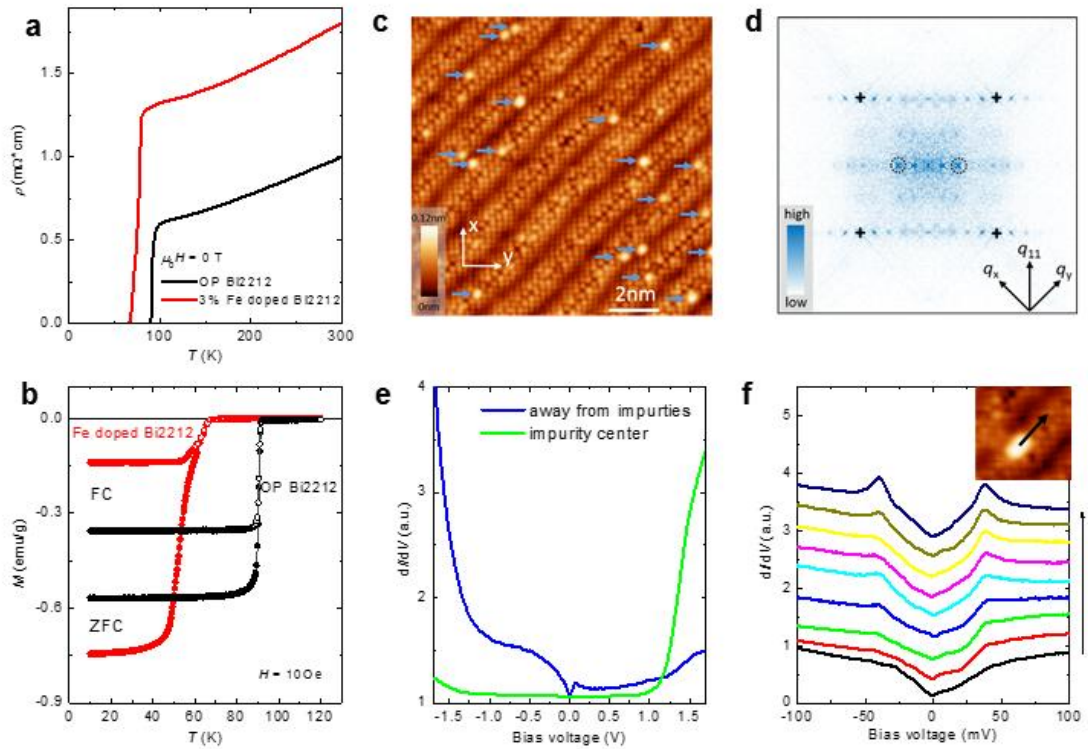

**Fig. 1 | Sample characterization and identification of iron impurities by using a tungsten STM tip. a**, Temperature dependent resistivity measured in 3% Fe-doped and optimally doped (OP) Bi2212 samples. **b**, Temperature dependence of magnetization measured in 3% Fe-doped and OP Bi2212 samples in the zero-field-cooling and field-cooling processes. **c**, Topographic image measured on the cleaved surface of 3% Fe-doped Bi2212 sample (setpoint conditions: $V_{bias}$ = 1.3 V, $I_t$ = 50 pA). The bright spots, which are indicated by arrows, are the locations of Fe impurities. **d**, Fourier transformed pattern to the topographic image in **c**. The crosses mark the positions of Bragg points, and the circles mark the characteristic spots derived from supermodulations along the (11) direction. **e**, Tunneling spectra in wide energy window measured at the Fe impurity site and away from any impurities ($V_{bias}$ = 1.4 V, $I_t$ = 100 pA). A considerable enhancement of differential conductance can be observed at bias voltages above 1 V on the spectrum measured at the Fe impurity site. Therefore, Fe impurities can be clearly visualized as bright spots in the topographic image measured with a bias voltage of 1.3 V. **f**, A series of tunneling spectra measured along the black line crossing an Fe impurity in the inset ($V_{bias}$ = −100 mV, $I_t$ = 100 pA). The coherence peaks are suppressed substantially at the Fe impurity site and recover gradually when the tip moves away from the impurity.

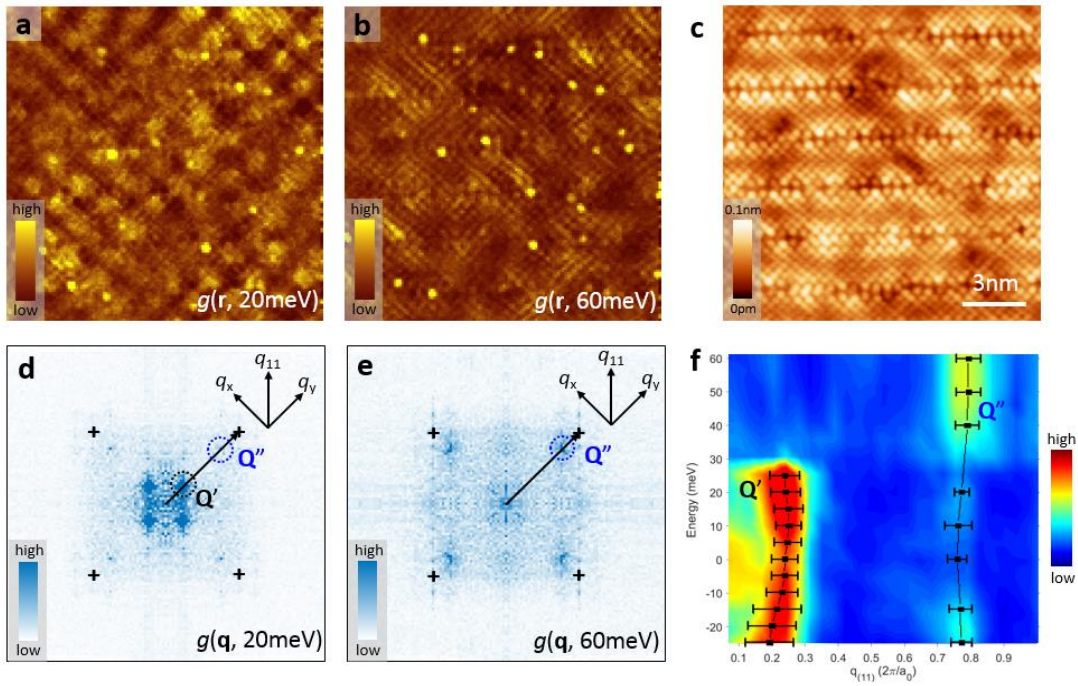

**Fig. 2 | DOS modulations in Fe doped Bi2212 measured by a tungsten tip. a,b**, QPI mappings measured at bias voltage of 20 and 60 mV, respectively ($V_{bias}$ = −100 mV, $I_t$ = 100 pA). Fe impurities can be identified in the QPI mapping measured at 60 mV. **c**, Topographic image measured in the same area of **a** or **b** ($V_{bias}$ = −100 mV, $I_t$ = 100 pA). Fe impurities do not show up because of the relatively low bias voltage. **d,e**, FT-QPI patterns based on QPI mappings shown in **b,c**, respectively. The characteristic DOS modulations with scattering wavevectors of **Q'** ≈ (±¼, 0) $2\pi/a_0$ and **Q"** ≈ (±¾, 0)$2\pi/a_0$ are marked by black and blue circles, respectively. **f**, Colour plot of energy-dependent line profile in FT-QPI patterns taken along the $q_{(11)}$ direction as marked by the arrows in **d** or **e**. Full squares denote peak centers of the characteristic charge modulations determined by fitting the line profile of data at different energies to Gaussian functions, and the error bars are the full widths at half maximum of the Gaussian functions. The feature of **Q'** spot mainly appears in the energy range of |E| < 30 meV while **Q"** spot only becomes pronounced at high energies above 35 meV, indicating two different origins for these two orders.

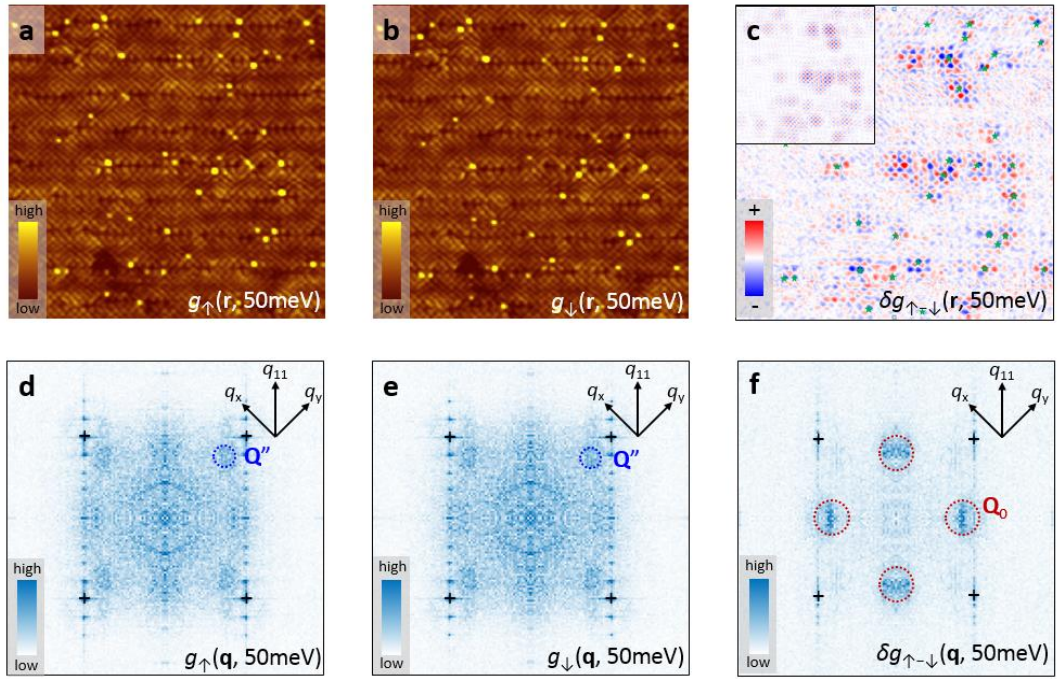

**Fig. 3 | Incommensurate AF order resolved by a spin polarized Cr tip. a,b**, QPI mappings measured at +50 mV and under 0 T by a Cr tip polarized by the magnetic fields of +1.2 and −1.2 T, respectively ($V_{bias}$ = −150 mV, $I_t$ = 100 pA). The positive (negative) sign of the magnetic field for polarizing the tip is denoted by '↑' ('↓'). **c**, Spin-difference LDOS mapping of $\delta g_{\uparrow-\downarrow}$(**r**, 50 meV) = $g_\uparrow$(**r**, 50 meV) − $g_\downarrow$(**r**, 50 meV). Clear sign-alternating periodic modulations can be observed near the Fe impurities (marked by stars). **d-f**, FT patterns to images in **a-c**, respectively. In **f**, four newly emergent scattering spots are marked by red circles with their center location wave vectors of about **Q**$_0$ = (±0.85π/$a_0$, ±0.85π/$a_0$). However, the scattering spots at **Q**" are absent in **f**. The inset in **c** shows the inverse Fourier transformed pattern of the spatial QPI taken on the FT-QPI data expressed by the complex values enclosed by the red circles in **f**.

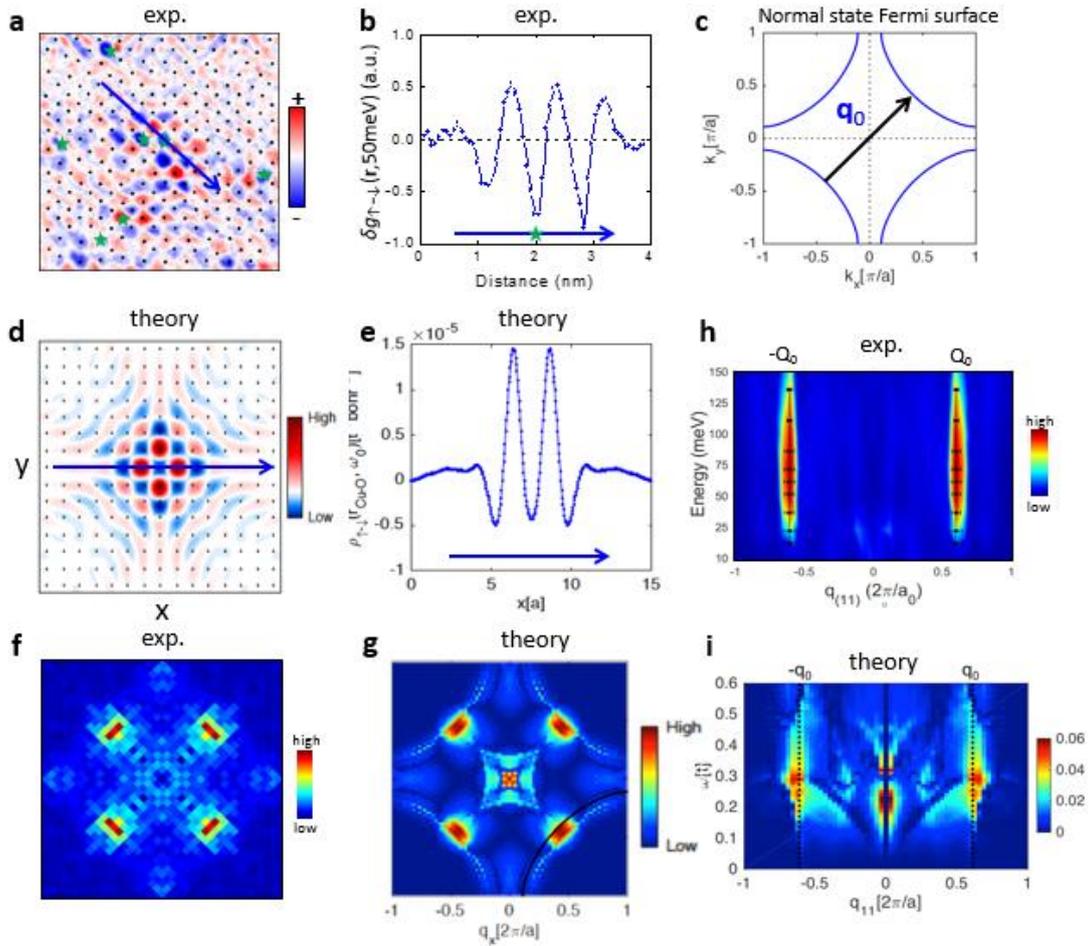

**Fig. 4 | Analysis of the emergent incommensurate AF order and theoretical explanation.**
**a**, $\delta g_{\uparrow-\downarrow}(r, 50\text{meV})$ measured in another area ($V_{bias}$ = −150 mV, $I_t$ = 100 pA). Bi sites (with Cu atoms below at exactly the same positions) derived from the topography are marked by black dots. The green stars mark the positions of the doped Fe atoms. **b**, Spatial dependence of $\delta g_{\uparrow-\downarrow}(r, 50\text{meV})$ measured along the blue arrowed line shown in **a**. **c**, Sketch of the normal-state Fermi surface showing a wave vector $q_0$ connecting two nodes. The length and the orientation of $q_0$ are close to $Q_0$ marked in Fig. 3f. **d**, A simulated spin-difference LDOS mapping at a bias of $\Delta_0$ (= $0.3t$) by using the single-band Hubbard model with one magnetic impurity. The black dots represent the Cu sites. **e**, The line cut of spin-difference LDOS along the arrowed line (Cu-O direction) shown in **d**. **f**,**g**, Experimental and simulated FT-QPI patterns by carrying out the Fourier transform to **a** and **d**, respectively. **h**,**i**, Energy dispersion of the antiferromagnetic order along the nodal direction in the spin-difference FT-QPI patterns from the experiment and the theoretical simulation, respectively.

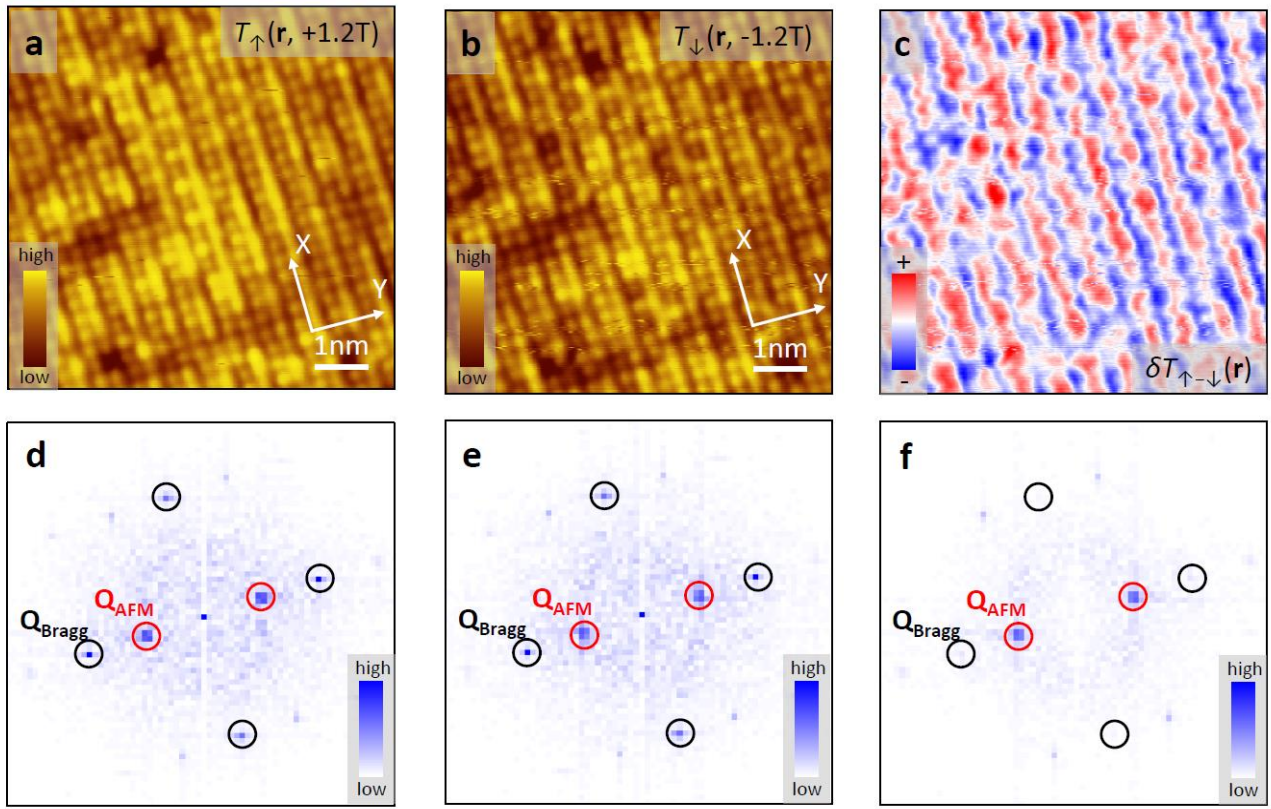

**Extended Data Fig. 1 | Characterization of spin-polarized tips on Fe$_{1+x}$Te. a-b**, the topographic image of Fe$_{1+x}$Te measured in the same region at the magnetic field of +1.2 T and -1.2 T using a spin-polarized Cr tip ($V_{bias}$ = 5 mV, $I_t$ = 100 pA). The sign of the magnetic field for polarizing the tip is denoted by '↑' or '↓'. **c**, The difference of the topographic images $\delta T_{\uparrow-\downarrow}(\mathbf{r})$ = $T_{\uparrow}(\mathbf{r}, +1.2T)$ - $T_{\downarrow}(\mathbf{r}, -1.2T)$ illustrating the scenario of the bi-collinear antiferromagnetic order in Fe$_{1+x}$Te. Here the background lattice structure information is subtracted out. **d-f**, FT patterns to images in **a-c**, the location of **Q**$_{Bragg}$ and **Q**$_{AF}$ are marked by the black and red circles.

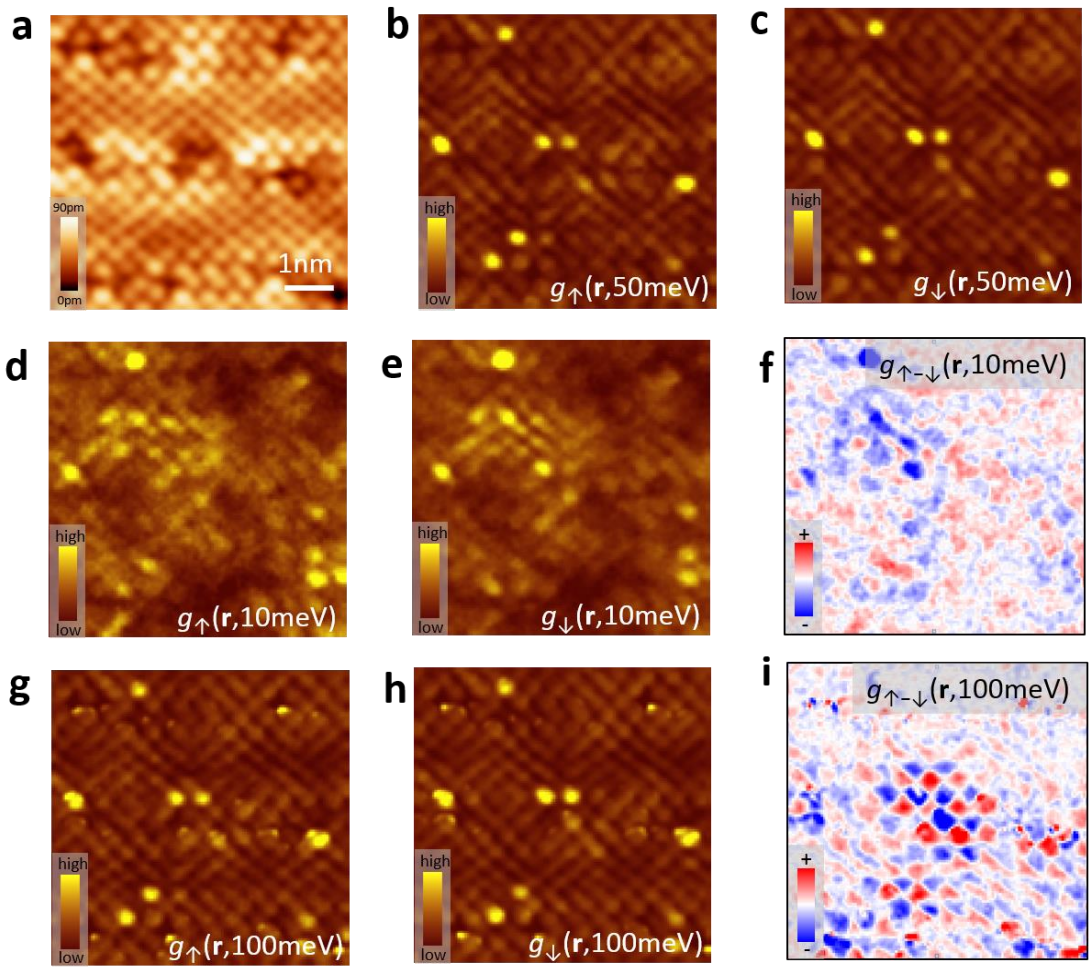

**Extended Data Fig. 2 | The incommensurate AF order measured at different energies.**
**a**, Topographic image measured by a Cr tip ($V_{bias}$ = −150 mV, $I_t$ = 100 pA). **b-e,g,h** QPI mappings measured at different energies and under 0 T by a Cr tip polarized under magnetic fields of +1.2 and −1.2 T ($V_{bias}$ = −150 mV, $I_t$ = 100 pA). These mappings are recorded in the same area of **a**. **f,i**, $\delta g_{\uparrow-\downarrow}(\mathbf{r},E)$ measured at different energies. $\delta g_{\uparrow-\downarrow}(\mathbf{r},E)$ measured at 50 meV is shown in Fig. 4a.

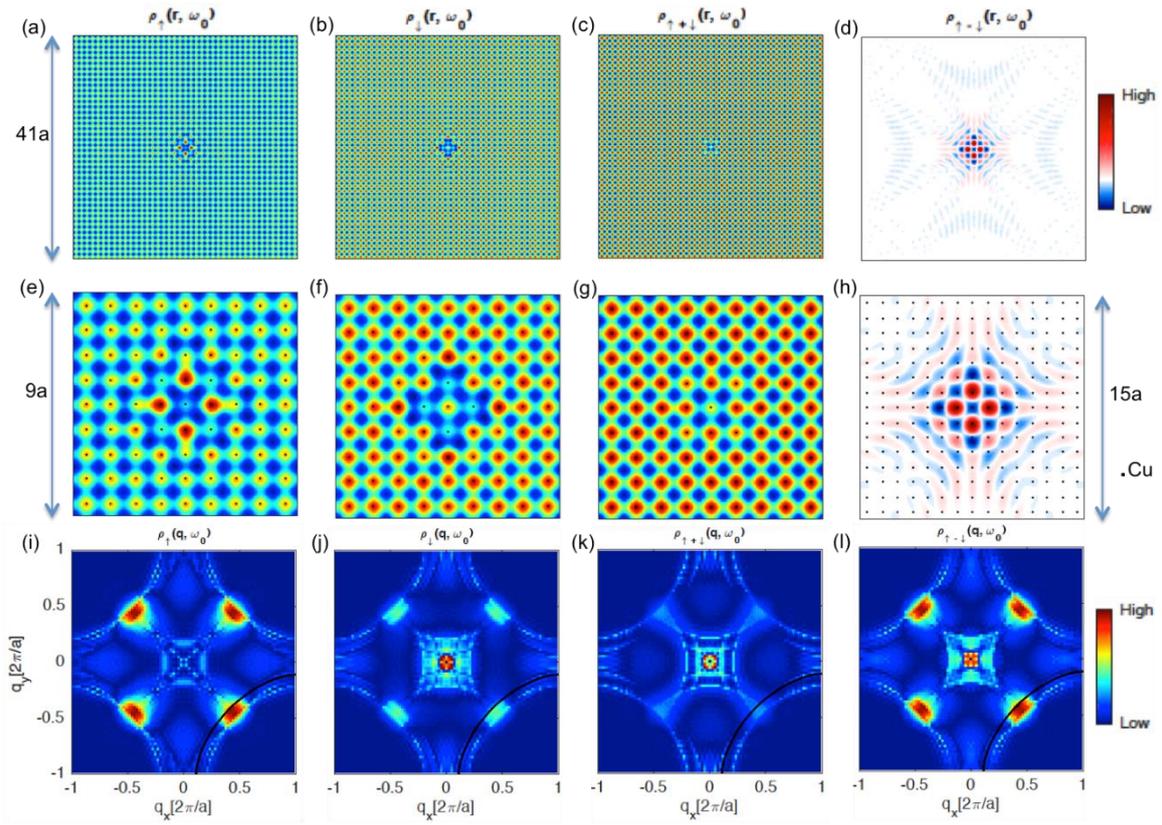

**Extended Data Fig. 3 | Calculated spin-resolved, spin-sum, and spin-difference LDOS maps at a bias of 1.25$\Delta_0$ (= 0.375$t$). a,b**, Spin-resolved LDOS maps over a large (41$a$ x 41$a$) area and **e,f**, corresponding zoomed-in views. **c,g**, Spin-sum LDOS maps and **d,h**, spin-difference LDOS maps in different scale. **i-l**, Spin-resolved, spin-sum, and spin-difference QPI obtained by Fourier transforming the corresponding LDOS maps are shown in **i-l**, respectively. In panel **h**, black dots represent Cu sites; and in panel **l**, the black curve traces a part of the normal-state Fermi surface.

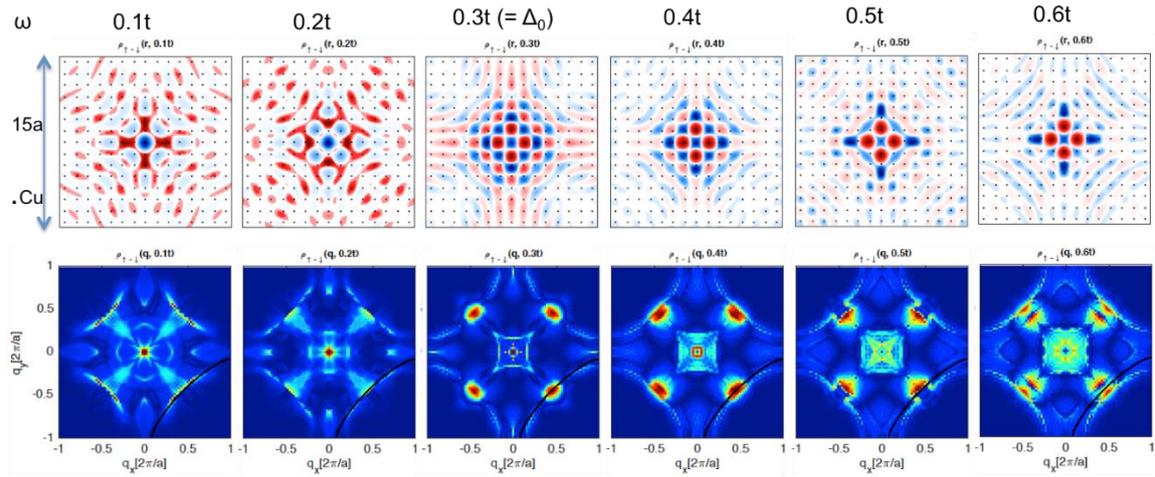

**Extended Data Fig. 4 | Calculated spin-difference LDOS maps (top panel) and corresponding QPI (bottom panel) at different energies and taking into account the AF correlations.** The energies are from $0.1t$ to $0.6t$ with an energy interval of $0.1t$ from left to right.

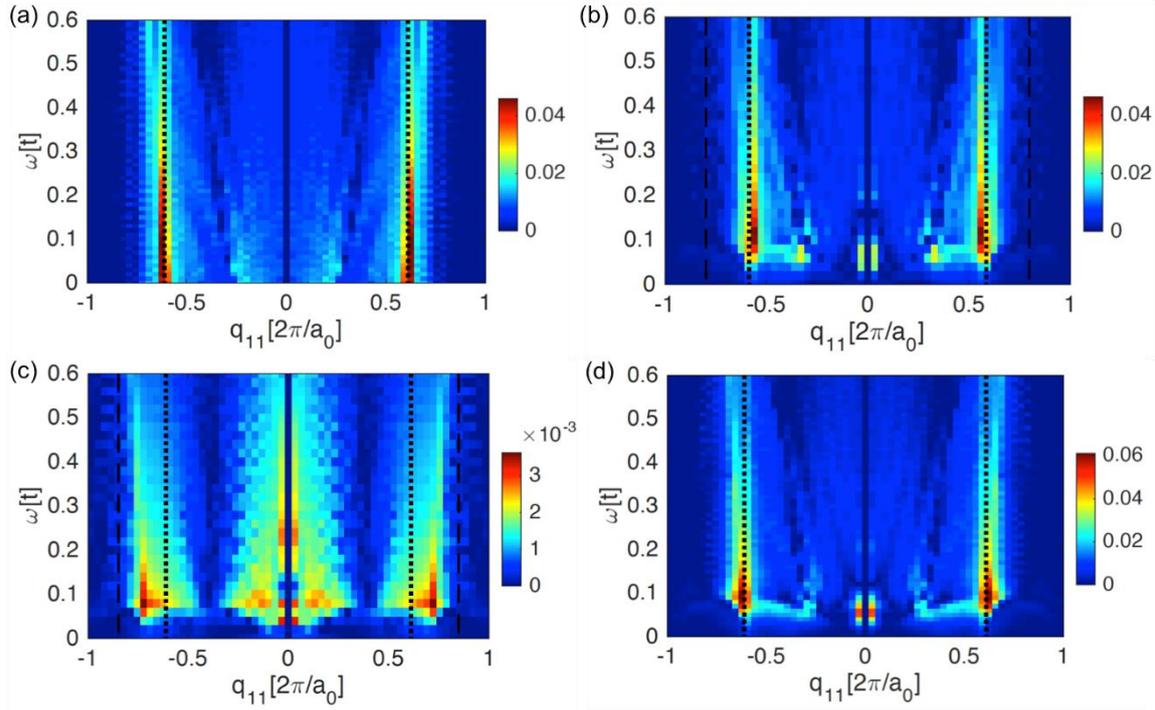

**Extended Data Fig. 5 | Calculated spin-difference QPI dispersion** for the normal-state including antiferromagnetic (AF) correlations (a); (b) and (c) show spin-difference QPI dispersion for the *d*-wave superconducting state with gap amplitude $\Delta_0 = 0.1t$ and NNN hopping $t' = -0.3t$ with and without taking into account AF correlations, respectively, (d) shows the same quantity for the *d*-wave superconductor with gap amplitude $\Delta_0 = 0.1t$ and $t' = -0.4t$ including AF correlations. Dotted black line represents the magnitude of the nodal nesting wavevector $\mathbf{q}_0$ in each case.

# Supplementary Information
# Ubiquitous Antiferromagnetic Order Visualized Directly by Spin-Polarized Tunneling Spectroscope in Optimally Doped $Bi_2Sr_2CaCu_2O_8$ Single Crystals with Magnetic Fe Impurities

**S1. Magnetization data and related analysis**

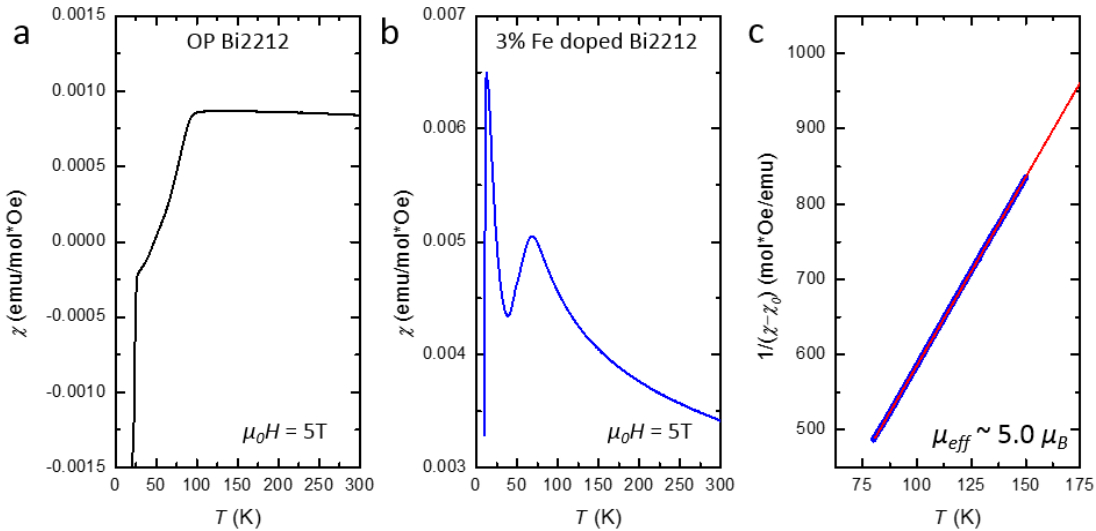

**Fig. S1 | Temperature dependent magnetization and fitting result by Curie-Weiss law. a,b**, Magnetic susceptibility of OP Bi2212 and 3% Fe-doped Bi2212 single crystal samples measured under 5 T. **c**, Temperature dependence of $1/(\chi-\chi_0)$ for the sample. The red line represents the linear fits of data according to the paramagnetic Curie-Weiss law. The slope gives $1/C_0$ and the intercept provides the value of $T_\theta/C_0$.

In order to study modulations induced by the Fe impurities doped in OP Bi2212, it is crucial to determine whether the Fe impurity is magnetic or non-magnetic. We measured the magnetization of OP Bi2212 and 3% Fe-doped Bi2212 under a high magnetic field of 5 T, as shown in Fig. S1. The enhancement of the magnetic susceptibility (compared to the OP

Bi2212) in Fig. S1b indicates the existence of local magnetic moment. In order to determine the magnetic moments induced by the Fe dopants, we fit the magnetic susceptibility by using the Curie-Weiss law

$$\chi = \chi_0 + \frac{C_0}{T + T_\theta}$$

where $C_0 = \frac{2zN_A\mu_B^2 p_{eff}^2}{3k_B}$. $2z$ is the molar concentration of Fe impurities and $p_{eff}$ is the local magnetic moment per Fe impurity site. We adjust the $\chi_0$ value to get temperature dependence of $1/(\chi - \chi_0)$ and then fit the data with a linear function. The slope of linear line gives $1/C_0$ and then we get the averaged magnetic momentum for each Fe site, which is $p_{eff}$ ~ 5.0. Thus, Fe dopants induce local magnetic moments when the doping level is 3% in our sample.

**S2. Another way to confirm the magnetic nature of Fe impurities**

We measure the topographic images on the 3% Fe-doped Bi2212 under zero magnetic field using Cr tips polarized by magnetic fields of +1.2 and -1.2T. These two images obtained in the same field of view are shown in Fig. S2a,b. The black circles in Fig. S2c help us to locate the positions of the Fe impurities in this area. In Fig. S2d, we show the difference of these two topographic images, namely $\delta T_{\uparrow-\downarrow}(\mathbf{r})$. An obvious distinction is observed nearby the Fe impurities, which is regarded as the consequence of the difference of spin 'up' and spin 'down' density of states nearby the Fe impurities. This would not happen if the Fe impurities were non-magnetic in nature, since in a *d*-wave superconducting condensate the bound states induced by non-magnetic impurities are spin degenerate. Thus the observed distinction between spin-up and spin-down tips provides a strong evidence that the Fe impurities are magnetic ones.

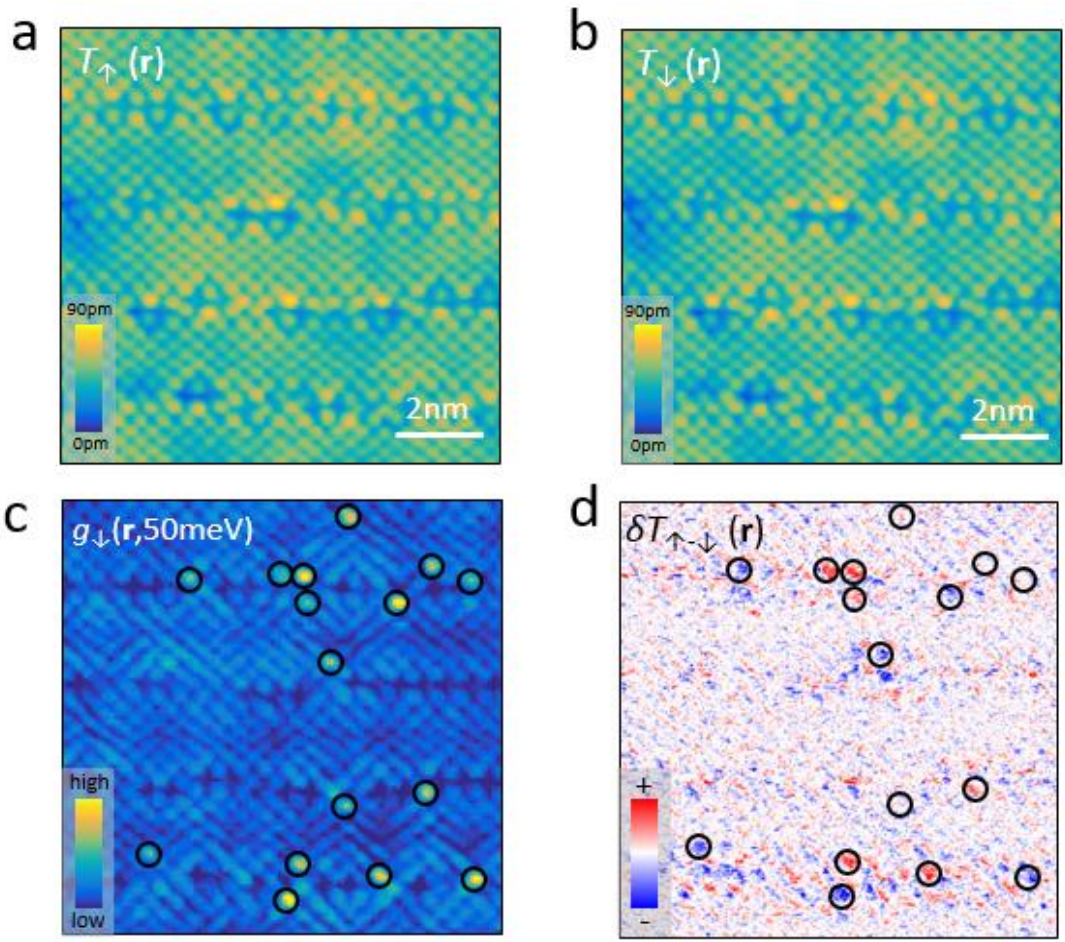

**Fig. S2 | Another way to confirm the magnetic nature of Fe impurities. a,b,** Topography of the 3% Fe-doped Bi2212 measured under zero magnetic field by the Cr tip polarized by magnetic fields of +1.2 and −1.2 T. **c,** The corresponding QPI mappings $g_↓(r,50meV)$ measured at +50 mV and under 0 T by a Cr tip polarized by magnetic fields of −1.2 T. The Fe impurities are marked by black circles. **d,** The difference of the topographic images $\delta T_{↑-↓}(r)$ = $T_↑(r)$ - $T_↓(r)$, which indicates that the spin-difference signals are mainly located near the impurities marked by black circles. The observation of the distinct spin-difference signals in the topographic images indicates the magnetic signature of the Fe impurities.

## S3. Analysis of iron impurity

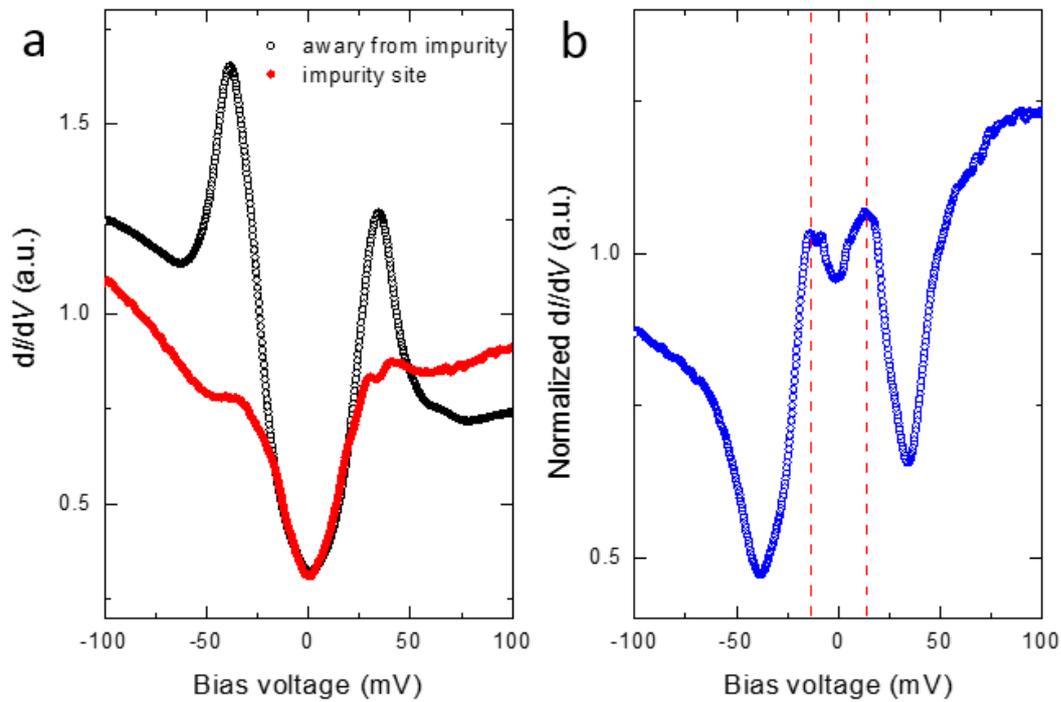

**Fig. S3 | Comparison of spectra measured at and away from the Fe impurities. a**, The averaged spectra taken at one impurity site (red solid circle) and away from the impurity (black open circle). ($V_{bias}$ = −100 mV, $I_t$ = 100 pA). The coherence peaks are clearly suppressed at the impurity site, as shown by the red solid circle curve. **b,** The divided result of the spectrum measured on the impurity site normalized by that measured away from the impurity. Two resonance peaks with additional subfeatures appear in the in-gap region, which are illustrated with two vertical lines at energies near 15meV.

As shown in Fig. S3, we compare the averaged spectrum taken at and away from one impurity site. The background (away from the Fe impurity) spectrum is the average of the spectra measured at locations of about 2nm away from the Fe impurity. Each spectrum is obtained through averaging over about 8 spectra. It is very clear that the coherence peaks are suppressed substantially near the impurities. Then we divide the averaged spectrum taken at the impurity by the one taken away from it. The normalized one shows two broad resonances peaked near ±15meV with some additional sub-features. Furthermore, it should be noted that the differential conductance on the impurity site is higher when the measured energy goes beyond 50meV. This allows us to identify the Fe impurities as bright spots on the differential conductance maps by using a high positive bias energy.

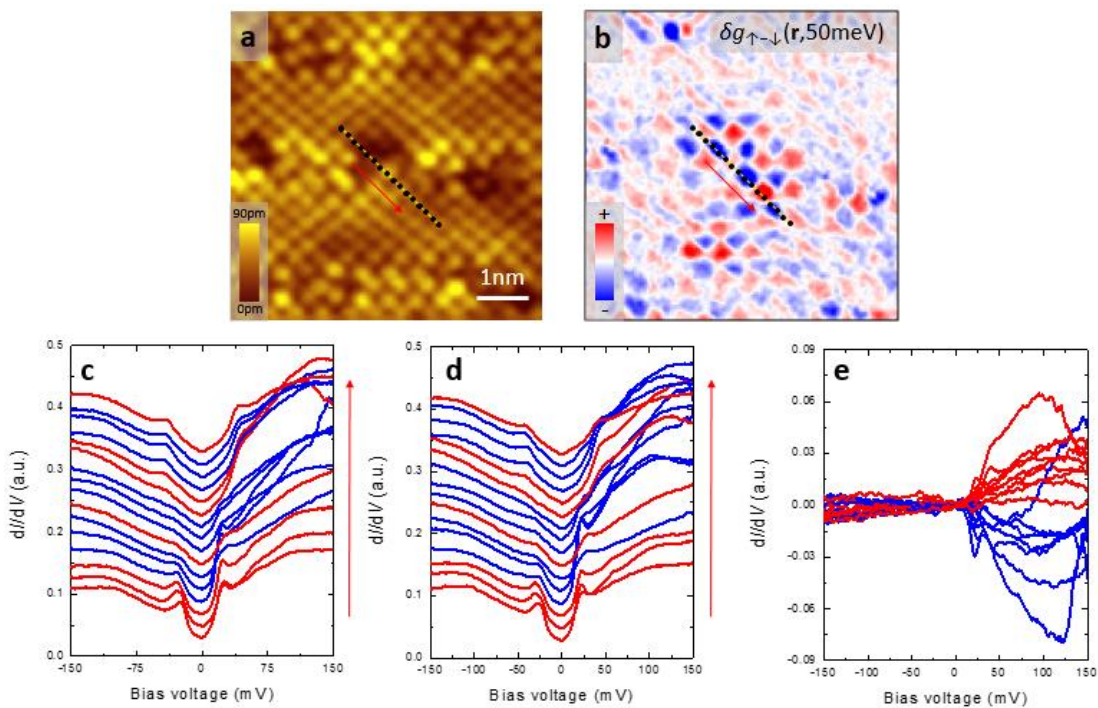

**Fig. S4 | Spin-resolved spectra measured by the Cr tip polarized by the positive and negative magnetic field. a,b**, The topographic image and the corresponding $\delta g_{\uparrow-\downarrow}(\mathbf{r}, 50\text{meV})$ ($V_{bias}$ = -150 mV, $I_t$ = 100 pA). **c,d**, spin-resolved spectra measured at the black dots along the red arrows in **a** and **b** using the Cr tip polarized by applied magnetic fields of +1.2T and -1.2T($V_{bias}$ = -150 mV, $I_t$ = 100 pA). The spectra shown in **c** and **d** which are shown as the red(blue) curves are measured at the position where $\delta g_{\uparrow-\downarrow}(\mathbf{r}, 50\text{meV})$ is positive(negative). **e**,

the difference of spin-resolved spectra measured by the Cr tip polarized by the positive and negative magnetic field. The difference mainly locates above the Fermi energy, indicating the spin-polarized signals mainly appear at the positive energies in the range from 25meV to 150meV. Furthermore, this result also supports the sign-alternating periodic modulations in the spin-difference LDOS mappings.

## S4. Theoretical modeling results

We model the normal state by setting $t'$ = -0.3 and hole doping 15%. This system shows a $(\pi,\pi)$ antiferromagnetic order at a critical value $U_c$ = 2.2. We take $U$ = 2 < $U_c$ to model a system with AF correlations but no long-range AF order. Further, taking NN attractive interaction $V_{nn}$ = -1 leads to a homogeneous d-wave superconducting state with gap amplitude $\Delta_0$ = 0.3. Finally, we take a weak magnetic impurity with exchange potential $J$ = 0.5 positioned in the center of the square lattice and solve the BdG equations self-consistently initializing it with random magnetization. Spatial variation of resulting mean-fields is shown in Fig. S5. The d-wave gap order parameter $\Delta_d = (\Delta_{i,i+\hat{x}} + \Delta_{i,i-\hat{x}} - \Delta_{i,i+\hat{y}} - \Delta_{i,i-\hat{y}})/4$, where $\hat{\tau}$ represents a NN site, and is suppressed at the impurity site and recovers to the homogeneous value far from it, see Fig. S5a. Hole density $\delta_i$ shows a damped oscillating behavior as expected see Fig. S5b. The most interesting feature is nucleation of AF puddle around the magnetic impurity as seen in the magnetization plot Fig. S5c. The AF puddle extends more along the diagonal direction and amplitude of the induced magnetization is of the same order as one would expect for the long-range AF state. A scan of lattice LDOS along the diagonal direction shows that the coherence peaks are suppressed at the impurity site and recover quickly as one move away, see Fig. 6. Since the impurity is week ($J$ < $W$/10, where $W$ is bandwidth), impurity-induced bound state only appears at the edge of the spectral gap, far from the chemical potential. In this regard the observed AF state is not necessarily a feature of the $d$-wave superconductors but the global instability of the systems towards an antiferromagnetic state.

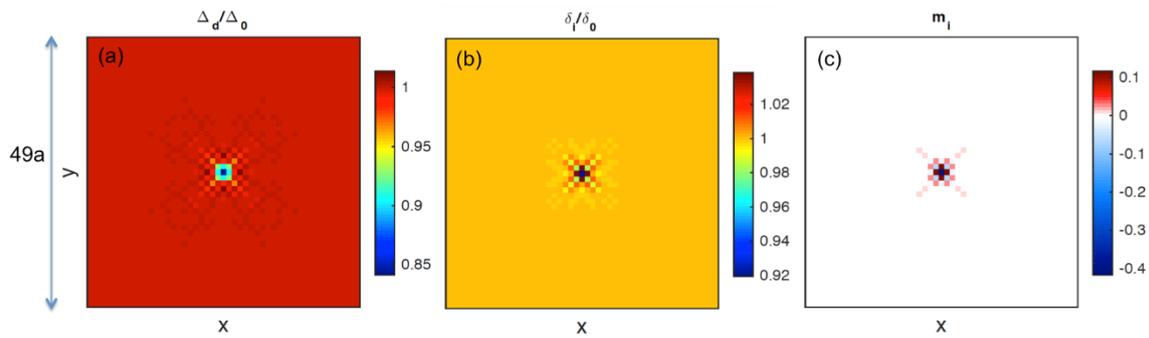

**Fig. S5 | Information of a magnetic impurity from theoretical modeling.** Spatial variation of (**a**) $d$-wave gap order parameter, (**b**) hole density per site, and (**c**) magnetization per site in a correlated $d$-wave superconductor (modeled by Hamiltonian in Methods part) with a magnetic impurity at the center of a $49a \times 49a$ lattice.

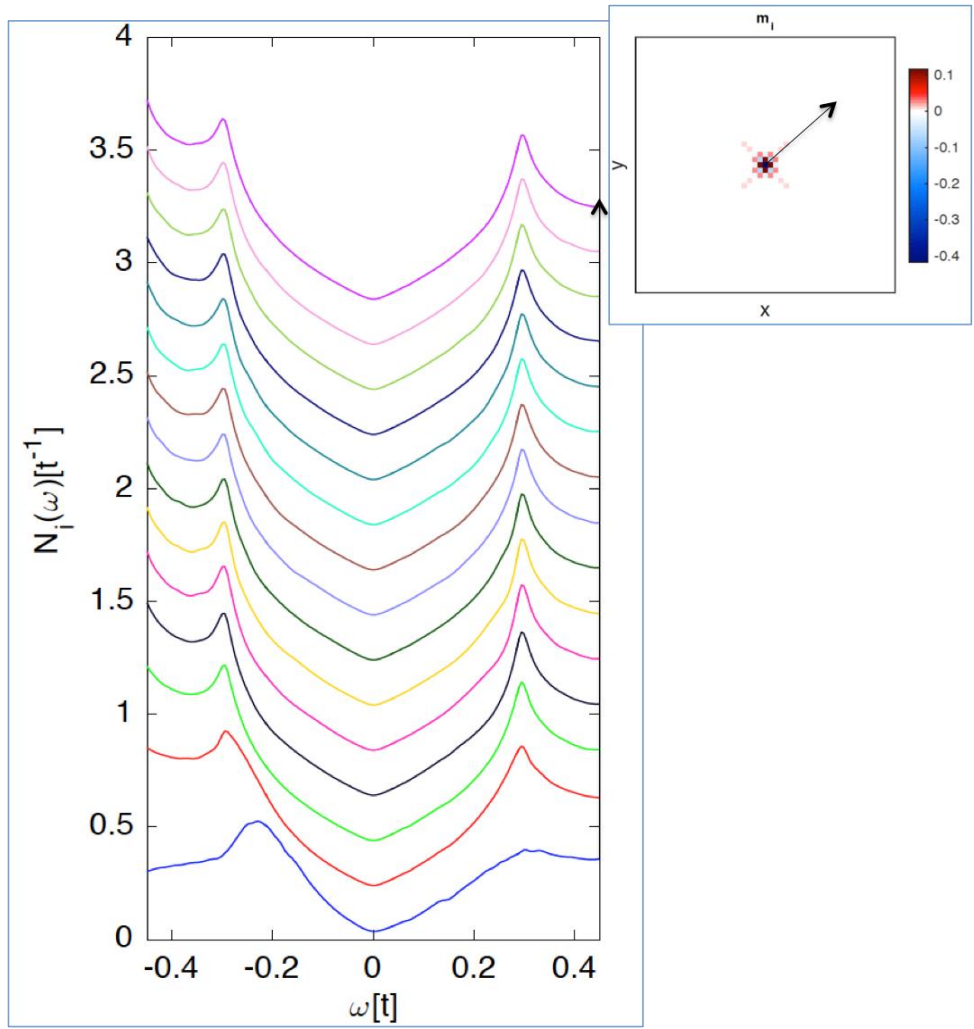

**Fig. S6 | Calculated scan of lattice LDOS in the diagonal direction away from the impurity**. The shown spectra are obtained along the arrow shown in the inset.